# Intensity Interferometry revival on the Côte d'Azur

Olivier Lai[a], William Guerin[b], Farrokh Vakili[a], Robin Kaiser[b], Jean Pierre Rivet[a], Mathilde Fouché[b], Guillaume Labeyrie[b], Julien Chabé[c], Clément Courde[c], Etienne Samain[c], David Vernet[d]

[a]Université Côte d'Azur, Observatoire de la Côte d'Azur, CNRS, Laboratoire Lagrange, France ;
[b]Université Côte d'Azur, CNRS, Institut de Physique de Nice, France ;
[c]Observatoire de la Côte d'Azur, CNRS, UMR Géoazur, France ;
[d] Observatoire de la Côte d'Azur, CNRS, UMS Galilée, France.

## ABSTRACT

Recent advances in photonics have revived the interest in intensity interferometry for astronomical applications. The success of amplitude interferometry in the early 1970s, which is now mature and producing spectacular astrophysical results (e.g. GRAVITY, MATISSE, CHARA, etc.), coupled with the limited sensitivity of intensity interferometry stalled any progress on this technique for the past 50 years. However, the precise control of the optical path difference in amplitude interferometry is constraining for very long baselines and at shorter wavelengths. Polarization measurements are also challenging in amplitude interferometry due to instrumental effects. The fortuitous presence of strong groups in astronomical interferometry and quantum optics at Université Côte d'Azur led to the development of a prototype experiment at Calern Observatory, allowing the measure of the temporal correlation $g^{(2)}(\tau, r=0)$ in 2016 and of the spatial correlation $g^{(2)}(r)$ in 2017 with a gain in sensitivity (normalized in observing time and collecting area) of a factor ~100 compared to Hanbury Brown and Twiss's original Narrabri Interferometer. We present possible ways to further develop this technique and point to. possible implementations on existing facilities, such as CTA, the VLTI ATs or the summit of Maunakea, which offer a unique scientific niche.

## 1. INTRODUCTION

Intensity interferometry was developed in the late 1950s by Hanbury Brown & Twiss in an impressive series of four papers [1, 2, 3, 4] and used throughout the 1960s, culminating in the measurement of the diameter of the 32 brightest stars in the Southern Hemisphere [5]. The principle of intensity interferometry is that the photon streams detected by separate detectors are correlated. This phenomenon goes right at the heart of the wave-particle duality of quantum mechanics. Indeed, the phenomenon can more readily be interpreted as intensity fluctuations along the wavefront, arising from interference between different frequencies of the light, as long as the source is unresolved. These correlations decrease as the source is partially resolved. It is much less intuitive to understand these correlations when considering the particle nature of photons, despite the fact that in practice it is precisely the arrival time of individual photons which provides the measure of the correlation.

### 1.1 Quantum optics

Consider a canonical Young's Double Slit experiment. It is often described that if one tries to detect the photon at either slit, the interference pattern disappears, because of the quantum nature of wave-particle duality; this is a consequence of what is known as Niels Bohr's Complementarity Principle. But this is an oversimplification, because if one were to place a photon counter at each slit and correlate the arrival times of photons, one would be able to predict whether the photons would interfere in the detector plane. The known separation of the slits, provides a solution to Schrödinger's equation, which is probabilistic; the uncertainty in arrival time (either instrumental or due to coherence), coupled with the Heisenberg uncertainty principle would determine which fringe the photon pair would contribute to. Leaving aside such speculative musings, we can also interpret the phenomenon of correlation of photon arrival time as the physical expression of Bose-Einstein statistics: At a given flux level, one would classically expect photons to arrive continuously and evenly, yet we know that photon (or shot) noise is the result of independent events that follow a random distribution. The correlation of photons can thus be interpreted as a measurement of the bunching that occurs in a beam of coherent particles following Bose-Einstein statistics.

These different interpretations are useful to examine the different physical processes that intensity interferometry can shed light on (pun noted, but not originally intended). Indeed, it is obvious that separate detectors with a variable spatial separation can be used to measure the decrease of coherence as the source gets partially resolved. However, different physical processes of light emission have different correlation functions, which can also be studied through temporal correlations.

## 1.2 Practical applications

The second order correlation function $g^{(2)}(\tau, r)$ is given by:

$$g^{(2)}(\tau,r) = \frac{\langle I(t+\tau,\rho+r)\ I(t,\rho) \rangle}{\langle I(t,\rho) \rangle^2}$$

As mentioned, measuring the correlation of photon arrival times at different spatial separations, $g^{(2)}(r)$, allows to measure the coherence of the light, which leads information on the visibility of the observed object. The temporal correlation contains information on the statistics underlying the emission process. For example, for fermions, $g^{(2)}(\tau=0)=0$, for thermal processes, $g^{(2)}(\tau=0)=2$, and for coherent emission (i.e. laser), $g^{(2)}(\tau)=1$ at all time, and intensity interferometry can therefore also be used to measure the spectral coherence and in particular the study of astrophysical lasers.

However, the major drawback of intensity interferometry is its sensitivity. Indeed, it is only photons in the same state (same wavelength, same polarization) that are correlated. If photons of different wavelengths are detected on the measuring device, they will be detected and correlated with other photons at the same wavelength but will not be correlated to photons at other wavelengths thereby decreasing the value of the observed correlation. However, fortuitously (or not, depending on one's perspective), broadening the band pass of detection does not alter the Signal to Noise ratio of the correlation: there are more photons and hence more correlations hence more signal, but in equal proportion to the increased noise. The total value of the number of photons increases, but the value of the observed correlation decreases. Also, photon counting devices usually have a dead time after a detection event and are insensitive while the diode resets. This electronic bandwidth was also, until recently a drawback for sensitivity, and imposed long integration times to be able to observe effects on faint astrophysical sources.

For reference, Hanbury Brown developed the Narrabri Intensity Interferometer in the 1960s [6]: two 6.5-meter telescopes could be arranged along a circular train track with a diameter of 188m. Each primary mirror was made of a mosaic of 252 hexagonal spherical mirrors, which were in no way cophased or coherenced, but simply focused the light on a large (42mm) photocathode at the prime focus of each light collector. With this device, at the shortest baseline possible (i.e. to perform the measurement at maximum visibility), they were able to obtain a SNR of 11.5 in 13.9 hours on $\alpha$ CMi, which is of magnitude 0.37 at V band. With these characteristics, it would take 6 years of continuous observing to measure the visibility of the brightest Active Galactic Nucleus (e.g. NGC 4151, at visible magnitude 9 [7]).

From the Calern single channel experiment (See Sect. 3), we find a gain of a factor ~100 in sensitivity with respect to Hanbury Brown original experiment. Although substantial, we are just at the dawn of possibilities, as single photon detector technology is continuously improving.

## 2. IMPROVING SENSITIVITY

As already noted by Hanbury Brown in 1974 [5], the most promising way to significantly improve the instrumental sensitivity of intensity interferometry is to spectrally disperse the light and measure correlations independently at different wavelengths using multiple photodetectors i.e., multiple spectral channels. Besides accessing valuable spectral lines to sound stellar surfaces and circumstellar structures, triggered by magnetic fields, mass loss or colliding winds, the signal-to-noise ratio (SNR) improves by averaging independent measurements as the square root of the spectral channels simultaneously correlated (assuming of course that the spectral channels all have the same spatial distribution, which is not always the case). Thus there are two technical issues facing the improvement of sensitivity: the first one is the development of very fast and sensitive Single Photon Avalanche Diodes (SPAD) *arrays*. The second issue is how to spectrally disperse the light into the multiple detection channels.

### 2.1 Detectors, APD arrays

Recent developments in Single Avalanche Photo Diode technology have produced commercially available components with unprecedented sensitivity, short dead time and high time resolution in the visible domain. The increase in time resolution of these detectors, coupled with modern Time to Digital Converters (TDC) which have count rates as high as several Mcps (millions of counts per second) allow for time resolutions of a few tens of picoseconds. This is smaller than a 50 to 100 picometer bandpass at visible wavelengths, so it gives an appreciation for how little of the light is actually being used in the correlation process. The combined dead time of the SPAD and the TDC are still limiting factors of a single channel's ultimate sensitivity

The development of arrays of such detectors is uncertain, dependent on the market needs. It is however critical to make any significant progress as the use of single SPAD per spectral channel quickly becomes prohibitive cost-wise and with diminishing returns due to the square root behavior of SNR. However, one dimensional arrays are more than adequate for our needs and may even be preferable if the output of a multi-mode fiber were to be dispersed using a simple prism, coupled with a lenslet array for each detector element. This would indeed be the simplest implementation of a multi-channel detector, although it should still be noted that 100 channels would only improve the SNR by a factor 10 (2.5 magnitudes). To become truly transformative, the number of simultaneous channels should number in the thousands or tens of thousands (e.g. 5000 channels increase the limiting magnitude by 4.5). The amount of data generated, or processing power if the correlations are computed on the fly are also not trivial.

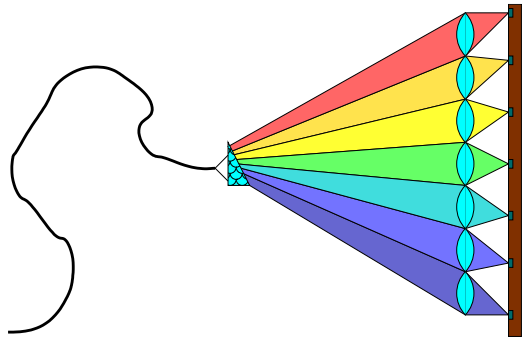

*Figure 1: Output of a multimode fiber onto a prism to disperse the light and refocusing it using a lenslet array onto an array of single photon avalanche diodes. Note that the output of the diode array has to be parallel and multiplexed into a TDC to be useful.*

### 2.2 Integrated optics spectral dispersion

Although it is possible to put the detector or detector array directly in the focal plan of the telescope, in practice, it is much more convenient to couple the light into an optical fiber and use the waveguide to bring the light to a delocalized detector, which can more conveniently be controlled in temperature and stray light. Multi-mode fibers have cores which can be as large a hundred microns or more, enabling very efficient coupling, and these cores are usually compatible with the photosensitive areas of SPADs. Although the output of such a fiber can be fed into a simple bulk optics spectrograph, the size of the fiber core, as well as the space requirements may make it much preferable to use single mode (SM) fibers which allow the use of fibered Fabry-Pérot etalons, fibered Bragg gratings or integrated (on-chip) spectrograph technology, to work on a large number of simultaneous narrow band spectral channels, which may help to keep the volume and design of a high dispersion spectrograph reasonable.

However, to efficiently couple the light into SM fibers, the telescopes must be equipped with adaptive optics (AO); this can be a problem for some implementation schemes (e.g. on Cherenkov telescopes, or if the telescopes are not already equipped with AO) and the optical throughput and the decreasing Strehl ratio at short wavelengths (where intensity interferometry offers an advantage over amplitude interferometry) can also negatively affect the sensitivity.

### 2.3 Adaptive Optics or Photonic lanterns?

There are alternatives, however. Photonic lanterns split a multi-mode fiber into a number of single mode fibers, which can then each be dispersed spectrally. The output of all the SM fibers can then be co-added before or after detection of each spectral channel. Co-adding before detection allows to reduce the number of detectors and is therefore preferable, especially if photonic lanterns can be used in reverse to mix the light of different single mode channels at the same wavelength, as shown in Figure 2. A drawback of this approach is that each single mode fiber channel needs its own spectrographic device. Mass replicated integrated optics components can provide a cost-effective solution to this problem but is not solved as of yet.

Another possibility is to line up the single mode fibers output of a photonic lantern at the entrance slit of a bulk optics single mode spectrograph. This would require a 2-dimensional SPAD array in the focal plane and is not necessarily optimal either.

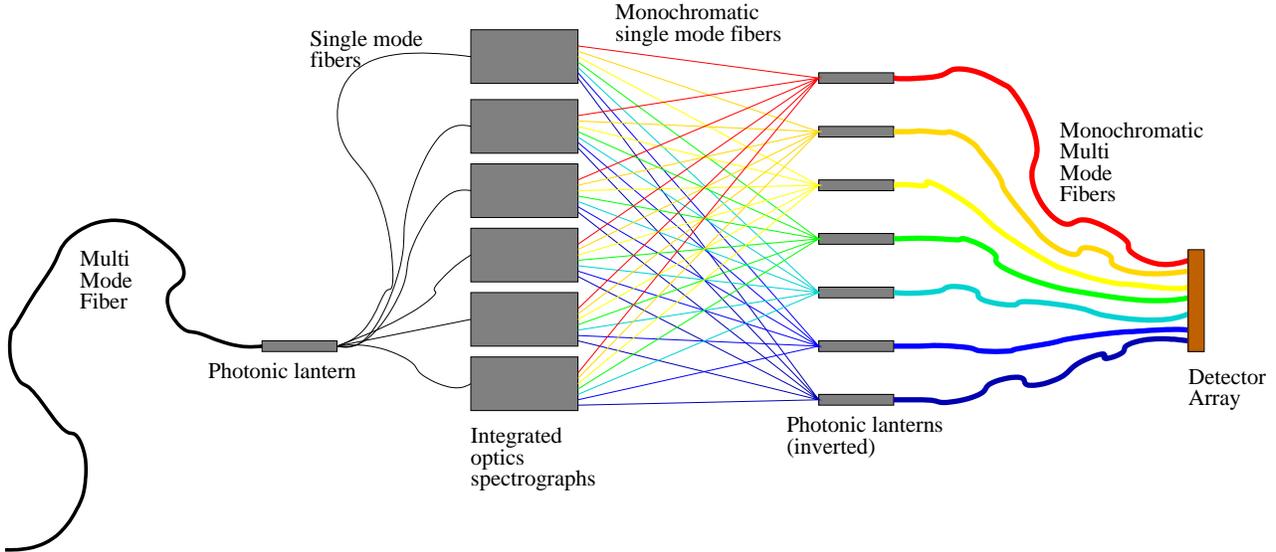

*Figure 2: Possible layout using integrated optics spectrographs and photonic lanterns to operate between multi-mode and single-mode fibers*

A trade study is needed in terms of overall throughput, efficiency and implementation/feasibility between using adaptive optics to couple the light directly into single mode fibers (which is hard at short wavelengths) or coupling into multi-mode fibers at the focus of the telescope and using photonic lanterns to split the light into multiple single fibers that can be coadded before or after detection or lined up on a spectrograph's entrance slit, although in the end, it will most likely be the detector technology that will provide the path forward.

## 3. RECENT EXPERIMENTS AT OBSERVATOIRE DE LA COTE D'AZUR

### 3.1 Temporal correlations

On three nights starting on February 20th, 2017, we obtained what we believe to be the first intensity correlation measurements with the light of a star (other than the sun, [8]) since HBT historical experiments [5]. The temporal photon bunching $g^{(2)}(\tau, r = 0)$, obtained in the photon counting regime, was measured for 3 bright stars, α Boo, α CMi, and β Gem [9]. The light was collected at the focal plane of the West telescope of the C2PU facility (Observatoire de la Côte d'Azur, OCA, Calern observing site, Figure ). The telescope provides a F/12.5 Cassegrain focus, but to improve the fiber coupling efficiency, we reduced the focal ratio to F/5.6 by inserting a focal reducer (using off-the-shelf components). In the focal plane, we placed our compact optical setup. It consisted of a dichroic beam splitter, a CCD guiding camera, a combination of filters and a multi-mode fiber (see Figure 3). The light reflected by the dichroic (λ < 650 nm) was used for guiding. The transmitted light was sent through a linear polarizer, then through two narrow band dielectric filters centered on $\lambda_0 = 780$ nm. The resulting bandwidth was 1 nm with an estimated throughput of 61% at the central wavelength.

The light was transported by a 20 m-long multimode graded-index fiber (MMF) with a core diameter of 100 μm, which was connected to a 50:50 fibered splitter such that two photodetectors could be used to compute the autocorrelation function without being limited by the dead time of the detectors, which is much larger than the coherence times (1nm of bandwidth corresponds to 500GHz, or 2 picoseconds of coherence time). Each output ports of the splitter were connected to a SPAD via two other MMFs of 1 m for the first detection channel and 2 m for the second. In order to avoid any spurious correlation induced by electronic cross-talk, we introduced an electronic delay between the two channels by using a 10 m shielded BNC cable for the second channel and 1 m for the first channel. The total delay (optical and electronic) between the two channels is $t_0 \simeq 45$ ns and was subtracted in the data processing. For each detected photon, the SPADs produce a 10 ns pulse, whose rising edge is detected and processed by a TDC with a time bin of 162 ps. The TDC was operated in the "Delay-Histogram" mode, which yields a good approximation of the intensity correlation function after proper normalization.

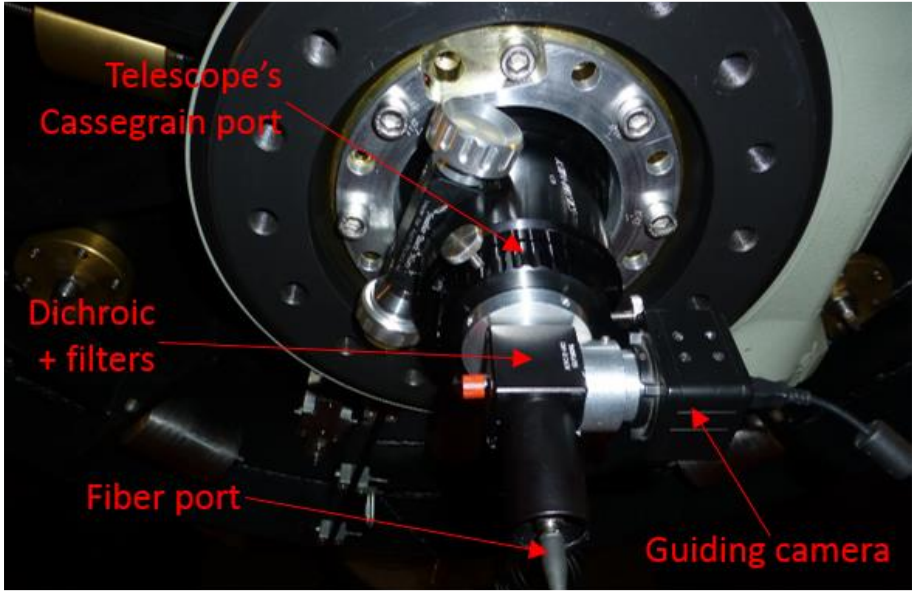

*Figure 3: Multi-mode fiber coupling device connected to the C2PU 1-meter telescope's Cassegrain port through a standard 2-inch eyepiece holder, in which two cascaded focal reducers are placed.*

For total exposure times of a few hours, we observed contrast values around $2\times10^{-3}$, in agreement with the theoretical expectation for chaotic sources, given the optical and electronic bandwidths of our setup. Results of $g^{(2)}(\tau)$ vs. $\tau$ for the three stars α Boo, α CMi, and β Gem are shown on Figure 4.

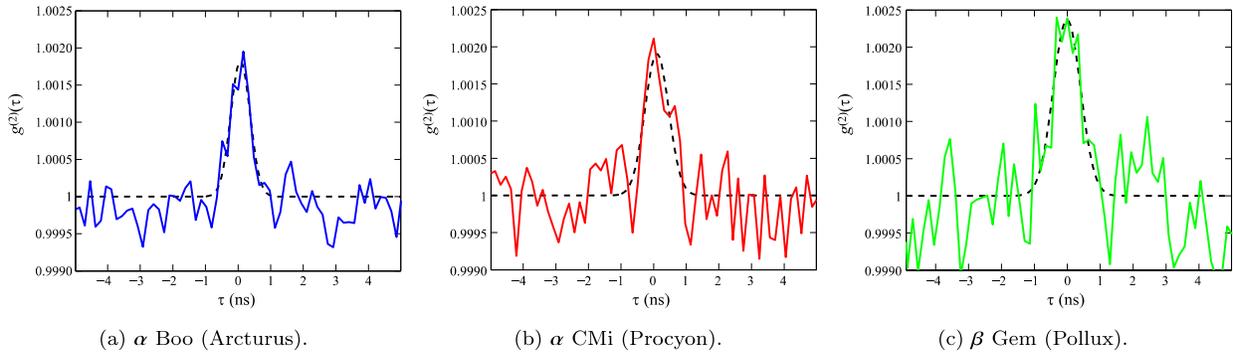

(a) $\alpha$ Boo (Arcturus).  (b) $\alpha$ CMi (Procyon).  (c) $\beta$ Gem (Pollux).

*Figure 4: Measured correlations of photon bunching on three bright stars measured on the 1 meter West telescope of C2PU. Cumulated integration times were 11h55' for α Boo, 4h35 for α CMi and 6h50' on β Gem.*

As we observed $\alpha$ CMi, we can carry out a direct comparison of sensitivity with the Narrabri interferometer: we obtained a SNR of 6.8 with only 4.6 hours observing time and with a collecting area approximately 85 times smaller (the single 1-meter telescope that we used had 9.7% central obscuration, compared to two 6.5m dishes). The gain in observing time is more or less made up by the decrease in SNR (a factor of roughly three in both cases), so our setup is about two orders of magnitudes more sensitive than the Narrabri Interferometer. The improvement in performance is due to the larger electronic bandwidth, the better quantum efficiency of the detectors, the fact that there are no correlation losses due to the separation of the telescopes (the minimal baseline of the Narrabri interferometer was 9.5 m) and our working wavelength of 780 nm, which is more adapted to the spectral type (F5IV) of α CMi than Narrabri's wavelength (443 nm). But the performance of our system could still be easily improved in two ways. First, a faster TDC and/or the choice of the "photon time tagging" mode, associated to an appropriate data post-processing system, would allow for improved measurements of the correlation function. Second, we could increase the fiber coupling efficiency by introducing a fast tip-tilt correction stage or low-order adaptive optics.

### 3.2 Spatial coherence

The presence of two independent but nearly identical 1 m telescopes separated by 15 m in the same facility (C2PU), made the evolution from temporal to spatial coherence relatively straightforward: the first successful observations were performed in October 2017 [10, 11]. For this, the fibered splitter used to feed both SPADs from the flux of a single telescope was removed and optical fibers coming from each telescope were connected to the two SPADs instead. Our fiber coupling system was duplicated (same guiding camera, same filters) and both replicas were installed in the focal planes of the two telescopes. Three bright stars (α Lyr, β Ori and α Aur) were observed for 4 nights, from 10th to 13th October 2017. The correlation peak emerged easily with the optical delay (connection wire lengths difference and time-dependent astronomical optical path difference) taken into account. The integration time on α Lyr was 11.1 hours spread over 3 nights and the on-sky projected baseline ranged from 9.6 m to 14.3 m, with a time-averaged value of 11.9 m. The contrast (Figure 5) was measured as $(0.89 \pm .09) \times 10^{-3}$ (1-σ uncertainty, i.e. SNR= 10). Taking into account the temporal resolution of the detection chain and a 11.9 m projected baseline, this value is quite consistent with the known value of the uniform disk angular diameter of α Lyr (3.198 mas) [12]. A more detailed study on other bright stars is presented in a detailed paper [11].

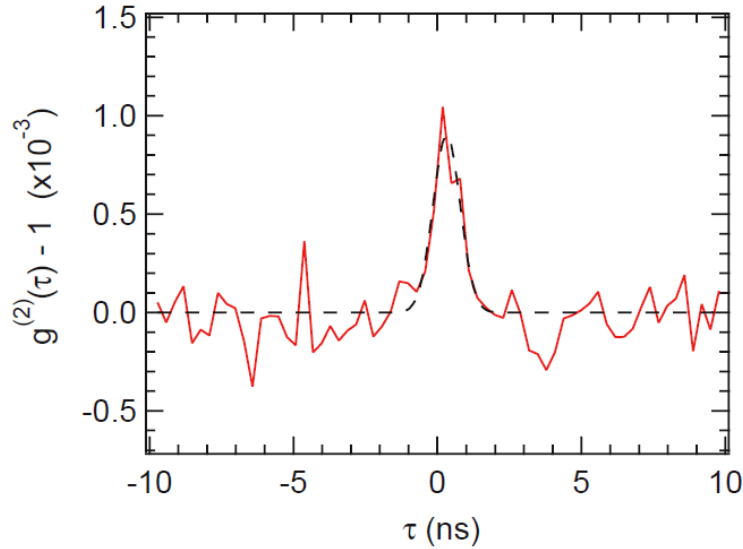

*Figure 5: Intensity correlation measured on Vega in October 2017, using the two C2PU 1m telescopes, with a projected baseline varying between 9 and 14.3m. The dashed line is Gaussian fit.*

### 3.3 Future plans at Calern

Following the straightforward success of intensity interferometry techniques on the C2PU facility at OCA, we have initiated a short-term plan to overcome its limitations and to improve its robustness for a further deployment on larger facilities in other observatories (see Sect. 4). The immediate plan however, is to improve our prototype experiment subsystem components to gain on the sensitivity (limiting coherent magnitude, i.e. source apparent brightness and effective visibility) whilst comparing its performances to operating optical amplitude interferometers like CHARA and/or NPOI through stellar sources with accurately known angular diameters.

A first step is based on introducing polarization beam-splitters to observe stars with extended atmospheres and/or wind envelopes simultaneously in two polarizations, parallel and perpendicular to the interferometric baseline. HBT's early observations at Narrabri on Rigel [13] have proven that substantial gains in accuracy on the visibility are required to measure any deviation from the uniform disk model. Since then, direct interferometers have been unsuccessful to achieve such measures (e.g. [14]) due to instrumental polarization effects (e.g. time-varying oblique reflections and cross-talk) introduced by the optical complexity of amplitude interferometers. Intensity interferometry is immune by construction to such effects since one can directly separate two polarizations and correlate them independently without any loss in SNR.

Moreover, a gain of √2 is expected for unpolarized sources, by co-adding the correlation functions of both polarization states.

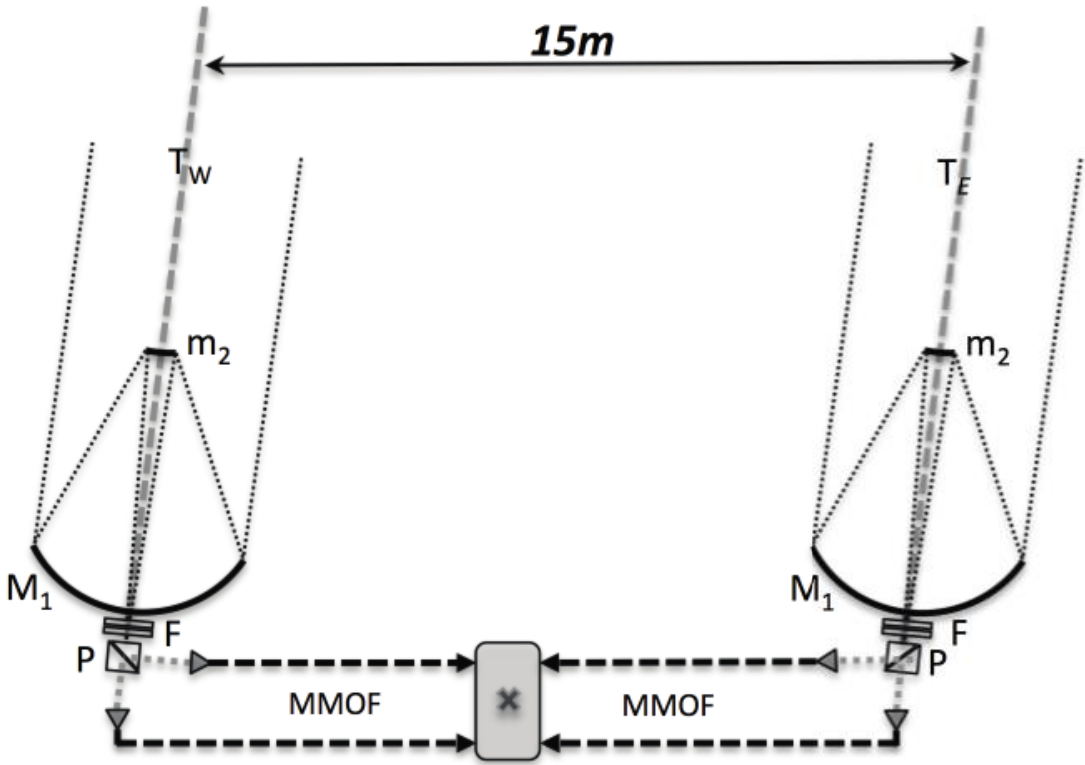

*Figure 6: Optical configuration of the I2C for stellar polarization and/or differential angular diameter measurements in continuum or emission lines such as Hα. Seeing limited images at the Cassegrain secondary foci are split in their two linearly polarized components and injected into two separate multimode optical fibers (MMOF), which transport the corresponding lights to SPADs. All four SPADs are connected to four input ports of the correlator X. Each polarization from both telescope is correlated to its counterpart from the other telescope so that $g^{(2)}(\tau,r)$ estimates are obtained simultaneously in polarized light parallel and perpendicular to the baseline. For C2PU the maximum East-West baseline reaches 15 m, corresponding to an angular resolution of 9.6 mas at 700 nm. F are 1 nm bandwidth filters either centered on the Hα line or at the 780 nm continuum to achieve differential visibility measurements.*

Figure 6 depicts the generic configuration of the I2C (Intensity Interferometer at Calern) project. This layout allows to observe differential effects between two spectral channels (one in the continuum and one on some emission lines such as Hα). This is interesting for hot stars, for instance Be or LBV like γ Cas or P Cyg, which have been formerly observed in amplitude interferometry at OCA [15].

A second step consists in increasing the angular resolution by performing intensity correlation experiments involving one or two telescopes from C2PU and the MéO (Métrologie Optique) Moon laser-ranging telescope [16], located at 150 m of the C2PU building on the Plateau de Calern. Synchronization of measurements between these distant telescopes can be achieved via cable or fiber connection with adequate control or measurement of transmission delays.

Most modern TDCs can be used in two modes: cross-correlation computation and photon time-tagging mode. Both modes can be used for astronomical intensity interferometry. If the electric signals of SPADs at the focus of telescopes, are fed into the different channels of the same TDC via coaxial cables, then the cross-correlation computation mode will deliver directly the photon arrival time correlation function with a time delay which needs to be computed and specified a priori (although in our case the data is obtained is 10 seconds bins, which can be adjusted before averaging the histograms and provide some leeway for a posteriori optimization). This operating mode requires a physical connection between telescopes through cables. On the contrary, if each telescope has its own SPAD, its own TDC (set in time-tagging mode), which delivers its time-tagging strings to its own computer, then no physical link is required between the telescopes. Correlations can be computed numerically *a-posteriori*. In this mode, all TDCs need to be accurately synchronized to a common time scale (e.g. UTC). This configuration is more expensive but more straightforward to extrapolate to very long baselines since no physical connection is required. Another advantage is that the correlation functions can be optimized numerically from the sets of photon arrival times, recorded independently by different TDCs and stored on independent mass storage devices. Contrary to correlations computed "on the fly" during the observation run, *a-posteriori* correlations computation allows for more elaborate algorithms, for example with time-varying and/or adjustable parameters. This configuration with no physical link between telescopes could be tested as a third step of our experiment. In our case, the necessary time reference for the TDCs would be provided by the local time standard of the MéO instrument. This time standard offers a temporal resolution and stability on the order of a few picoseconds, which is significantly better than the coherence length of our intensity correlation measurements (of the order of 3 cm, because of the time-resolution of our SPADs).

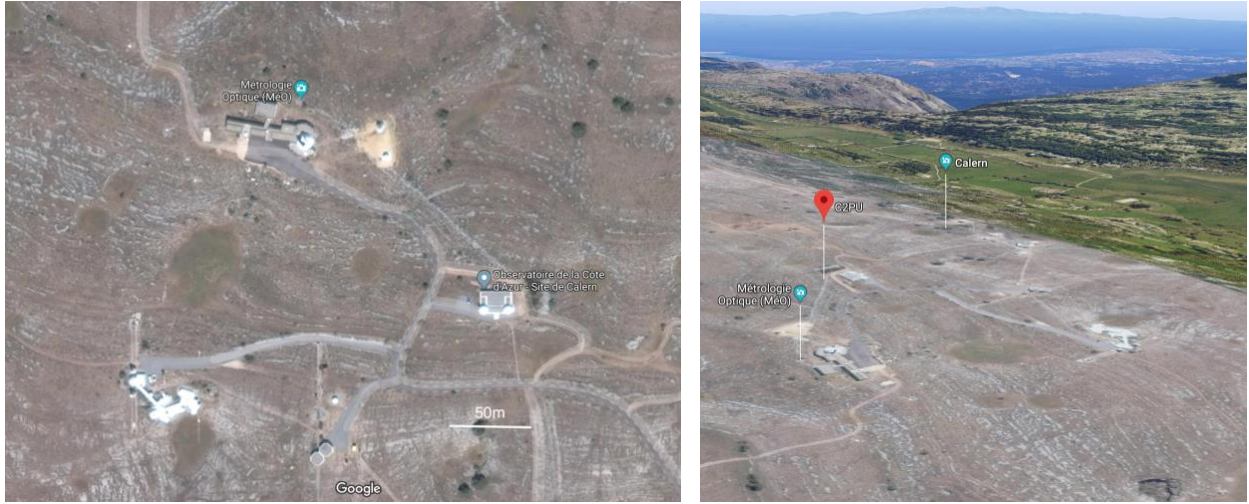

*Figure 7: The Plateau de Calern Observatory, with the two 1m C2PU telescopes separated by 15m on the right, and the MéO laser ranging 1.5m telescope 150 meters away to the North-West. The GI2T interferometer is visible at the bottom left of left view. Corsica is visible on the horizon of the view on the right.*

The layout of the 3-telescope intensity correlation experiments between the two C2PU 1 m telescopes and the MéO laser-ranging 1.5 m telescope is shown on Figure 7. The West-North-West direction of MéO with respect to C2PU offers much longer projected baselines than C2PU itself. They range between a few tens of meters and 150 m depending on the target's hour angle. This leads to sub-mas resolutions for visible wavelengths (e.g. in the range of 600-800 nm). Indeed, triple correlations between 3 telescopes would provide the closure phase quantity [17] but at the expense of a much higher exposure time for a given SNR. Several hundred hours of integration would be necessary with the OCA telescopes to attain the SNR value required to access any measure of phase information from the triple correlation quantity. However, the power of intensity interferometry, despite its intrinsic low SNR and sensitivity, lies in the fact that the correlation signal can be integrated over many nights since no fringe detection or tracking is necessary as in amplitude interferometry.

## 4. IMPLEMENTATION ON EXISTING FACILITIES

Intensity Interferometry is relatively easy to implement even in sites not initially designed for coupling of telescopes since no physical link is required, apart from a common time reference. We examine three possibilities for implementing intensity interferometry on existing sites.

### 4.1 Cherenkov Arrays

The future Cherenkov Telescope Array (CTA) will consist of tens of 4m to 23m telescopes spread over kilometric-sized areas. These characteristics would offer an unprecedented photon collecting area with sub-milliarcsecond angular resolutions and dense *(u, v)* plane coverage. Furthermore, the array can only be used for high energy physics purposes when the night sky is sufficiently dark, leaving an opportunity in bright time. Unfortunately, these telescopes are not designed to be diffraction-limited and have small aperture ratios. This makes the coupling into optical fibers or SPADs difficult. It may be possible to use one of their photomultiplier tubes, as the frame rate for the CTA cameras is 1GHz (1 ns time resolution but could potentially be increased if only one detector is used?). The field of each pixel is between 6' and 12', so an aperture and a narrow band filter would still need to be implemented. There may also be other simple solutions that we have not considered, such as optical cones or some sort of focal reducer optics to rescale the image. Nonetheless, the coupling issue is the most challenging aspect of implementing intensity interferometry on Cherenkov telescopes.

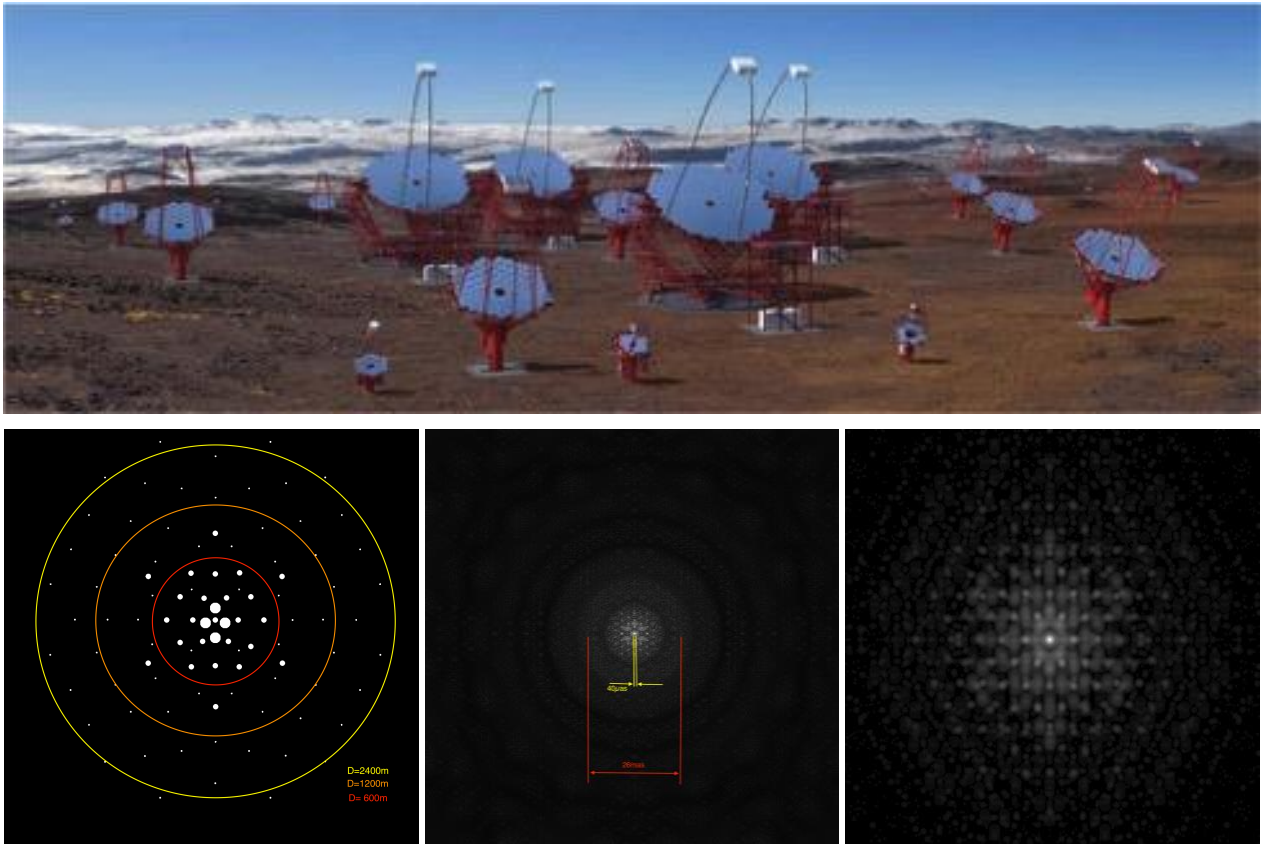

*Figure 8: Top, Artist's conception of the planned CTA-South array in the Atacama Desert (image credit: Gabriel Pérez Diaz, IAC). Left: position and relative size of the telescopes. The associated PSF is shown in the middle panel while the* u-v *plane coverage is shown on the right.*

Nevertheless, the collecting area of such arrays is extremely compelling: the largest telescopes are 24m in diameter, which is a collecting area of 415m$^2$, (there are four, so 1661m$^2$), The medium size telescopes are 12m in diameter, so an area of 113m$^2$ (times 25 so 2830m$^2$) and the "small" ones are 4m in size, so an area of 12m$^2$ x 70 = 125m$^2$. It is however not trivial to compute an effective limiting magnitude because of two complex effects: The first one is that when combining the light from telescopes of different sizes, the sensitivity is equal to that of an effective diameter equal to the geometric mean of both diameters. The second one is that the flux of redundant baselines can be co-added with no loss of coherence; this means that reconstruction schemes with multiple correlations of different co-additions can improve the estimation of the visibility compared with simple pairwise correlations; optimal reconstruction schemes may need to be borrowed from radio-astronomy. Despite all these difficulties, it is worth remembering that with a largest baseline of 2400m, Intensity interferometry on CTA would enable resolutions on the order of 40 µ-arcseconds.

### 4.2 Existing optical interferometers

Existing interferometers were designed for the purpose of amplitude interferometry and don't have much to gain in investing large amounts of time in intensity interferometry. However, an opportunity may exist on the ESO VLTI, when all the existing delay lines are being used with GRAVITY or MATISSE: The ATs, which will soon be equipped with adaptive optics, will not be useable for amplitude interferometry, but thanks to the ease of implementation, could easily be used for intensity interferometry with its own instrumentation. Integrations on objects need not be contiguous in time, and over time, signal could be accumulated. Preliminary contact with ESO was initiated by our team members.

### 4.3 OHANA on Maunakea

The OHANA (Optical Hawaiian Array for Nanoradian Astronomy) was an attempt to connect all the large telescopes on Maunakea into an amplitude optical/infrared interferometer using single mode fibers [18, 19, 20]. Four 8-meter telescopes (Gemini, Subaru, and both Keck 10m telescopes) as well as four 2-4m class telescopes (UKIRT, CFHT, IRTF, UH88") serendipitously arranged in a semi-circle with 800m diameter provide excellent $u$-$v$ coverage. Fringes were obtained between both Keck telescopes using single mode fibers in 2005 [19], but subsequent attempts were thwarted by unbelievably bad luck with weather. Attempts to connect CFHT with Gemini revealed difficulties of implementing amplitude interferometry in sites not originally designed for such. Despite the existence of underground pipes to carry power and network across the summit area, it would have been difficult to add our single mode optical fiber bundles to the existing routing and would have added much extra fiber length. For CFHT-Gemini, we installed our own pipe, which is now used for the GRACES project. Also, vibrations on facilities not designed for interferometry (and others) are often so large as to make interferometry challenging if not impossible. Active metrology is required at the very least.

Intensity Interferometry would be simpler to implement on such a site and offer amazing characteristics at short (visible) wavelengths: the largest baseline between Gemini and Subaru is 800 meters, which would yield a resolution of 100 $\mu$-arcseconds at 440nm. When TMT is eventually built on Maunakea with its 700m$^2$ collecting area, the baselines with Keck will be 1km and with Gemini up to 1650m. Of course, it will be difficult to obtain large amounts of telescope time with such telescopes, but the potential for high angular resolution is such that it is worth pursuing.

In the meantime, if we simply extrapolate the numbers from the Calern setup (single spectral channel) to a thought experiment on Maunakea using IRTF-CFHT-UH88", we find that we would be able measure the correlation of an unresolved source of magnitude 4 with a SNR of 5 in 1 night. IRTF-CFHT-UH88" form a right-angle triangle with ~300m baselines; this is a resolution of 500 $\mu$-arcseconds at H$\alpha$!

Including all the telescopes from the original OHANA array (namely both Keck telescopes, Gemini and Subaru) would increase the collecting area tenfold, and hence improve the limiting magnitude to 6.5.

### 4.4 Other future possibilities

It is worth mentioning a few other very interesting possibilities where large telescopes provide serendipitous baselines. Gemini South is 420 meters away from SOAR. Members of our team have contacted Brazilian astronomers who have access to both facilities, to discuss the interest of such measurements. Also, as already mentioned the TMT will be within a few kilometers range of the 8-meter telescopes on top of Maunakea. Similarly, the Giant Magellan Telescope (GMT) will be 1.7 kilometer from the two 6.5-meter Magellan telescopes, which corresponds to a resolution of 60 $\mu$-arcseconds at 500nm. Finally, we note that the ELT site will be 21km away from the VLT site, which provides a resolution of 5 $\mu$-arcseconds; Sirius-b, the closest known White Dwarf is expected to measure around $10^4$ km in diameter, which at a distance of 8.6 light years, is an angular size of a few tens of $\mu$-arcseconds.

## 5. FUTURE PROSPECTS

We are pursuing an active program to revive intensity interferometry at Université Côte d'Azur by joining forces between the quantum optics laboratory at Institut de Physique de Nice (INPHYNI) and Laboratoire Lagrange at OCA. Access to the two 1-meter telescopes separated by 15 meters and the 1.5 MéO telescope 150 meters away offers a perfect laboratory to develop solutions in real observing environments and is a good use of this class of telescopes. However, due to the sensitivity of Intensity Interferometry, access to large apertures will become necessary. As its main strengths are short wavelengths and long baselines, i.e. extremely high angular resolution, the types of targets that will be the niche of II will be compact objects with intrinsically high surface brightness, so highly energetic types of objects.

### 5.1 Original astrophysical applications

To make astrophysically meaningful measurements on compact, exotic objects, access to larger collecting areas becomes necessary. Scientific applications at short wavelength (and high surface brightness/energy density, given the resolution!) will be favored, and seem to point to compact objects such as white dwarfs, X-ray binaries. As mentioned in Section 4.4, White dwarfs are expected to be on the order of tens of micro-arcseconds and smaller, only accessible with baselines on the order of 10 kilometers. We can get a rough estimate of sensitivity of large arrays coupled interferometrically if simply assume as a first approximation that the SNR is proportional to the collecting area, although this is not strictly true due to the geometrical ratio of different size telescopes as well as redundant baselines co-additions. Under this assumption we then use our Calern results and normalize them for a SNR of 5 in one night of observing, and optimistically assuming 5000 spectral channels. The Ohana array would have a limiting magnitude of 11.2 and provide resolution between 0.25mas and 0.5mas (for 400nm to 900nm respectively). Adding the TMT would increase the limiting magnitude to 13.1 and with a resolution of 40 micro-arcseconds. Similar resolutions would be achieved by the GMT coupled with the Magellan telescopes, with a limiting magnitude of 12.3. Finally, the VLTs and the ELT, would provide a limiting magnitude of 13.6 with resolution 4 micro-arcseconds. These numbers are plotted on Figure 9. On the same figure, we have plotted the estimated diameter and know brightness of Sirius b, as well as the extrapolation for further White Dwarfs using their distance modulus. We have also plotted the estimated size and brightness of the accretion disk of NGC 4151 (and its distance modulus extrapolation). Clearly, intensity interferometry will be most useful on very bright surface brightness objects.

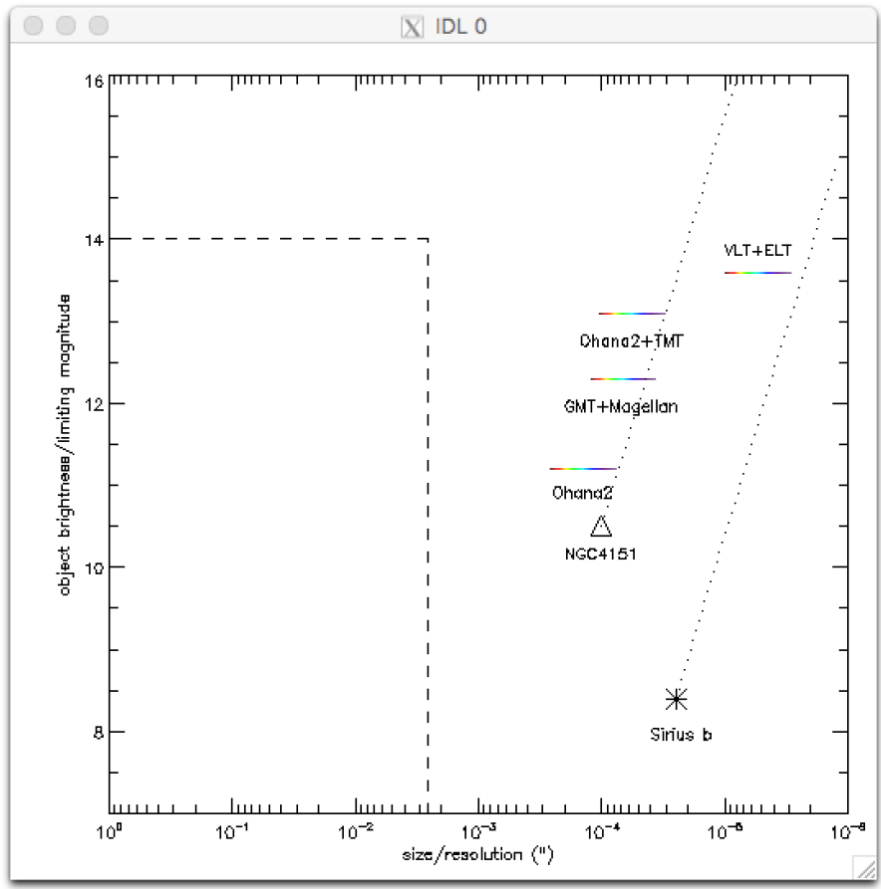

*Figure 9: Limiting magnitude and resolution for various arrays extrapolated from our Calern experiment, assuming a SNR of 5, one night of observing and 5000 spectral channels. Sirius b's size and brightness are also plotted and would be partially resolved with any of the kilometric arrays, as would the accretion disk (assumed to be on the order of $10^{-2}$pc) of NGC 4151.*

Other exotic objects that may be bright and compact enough would include X-ray binaries and associated accretion disks. Extragalactic accretion disks may also be observable: the accretion disk of NGC 4151, assuming a size of $10^{-2}$pc at 20Mpc, would have an angular size of 100 micro-arcseconds; its magnitude is around 9 in K band, but closer to 11 or 12 in V band. Brown dwarfs would also be easy to resolve but their very red spectral characteristics make them less well suited for intensity interferometry in the visible domain. Single Photon Infrared Avalanche Diodes would be needed for such observations.

ACKNOWLEDGEMENTS

We thank Antoine Dussaux for important contributions to this project. This work is supported by the UCA-JEDI project ANR-15-IDEX-01.